\documentclass[a4paper]{ISOStyle}
\usepackage{epsfig}

\begin{document}

\title{ISO and the Cosmic Infrared Background}

\author{Herv\'e Dole\inst{1}} 
  
\institute{$^1$ SIRTF/MIPS Team, Steward Observatory, University of
Arizona, 933 N Cherry Ave, Tucson, AZ, 85721, USA\\
http://lully.as.arizona.edu}

\maketitle 

\begin{abstract}
ISO observed, for the first time to such a high sensitivity level, the mid-
and far- infrared universe. A Number of deep surveys were performed to probe
the cosmological evolution of galaxies. In this review, I discuss
and summarize results of mid-infrared ISOCAM and far-infrared
ISOPHOT surveys, and show how our vision of the extragalactic infrared
universe has become more accurate. In particular, ISO allowed us to
resolve into sources a significant fraction of the Cosmic Infrared
Background (CIB) in the mid-infrared, and to probe a fainter
population in the far-infrared with the detection of the CIB
fluctuations. Together with other wavelength data sets, the nature of ISO
galaxies is now in the process of being understood.

I also show that the high quality of the ISO data put strong
constraints on the scenarios of galaxy evolution. This induced a burst
in the development of models, yielding to a more coherent picture
of galaxy evolution. 

I finally emphasize the potential of the ISO data archive in the field
of observational cosmology, and describe the next steps, in particular the
forthcoming cosmological surveys to be carried out by SIRTF. 

\end{abstract}

\section{INTRODUCTION}
The Cosmic Infrared Background (CIB) (e.g. Puget et al, 1996; Hauser
\& Dwek, 2001) is the relic emission of galaxy formation and
evolution, being composed of light radiated from galaxies since their
formation. The aim of the cosmological surveys is to resolve the CIB into
sources, and to provide data on galaxies at various wavelengths,
redshifts, with enough quality on individual objects as well as with
statistical significant samples, in order to address the questions of
the processes of galaxy formation and evolution, and the nature of the
galaxies. 

The CIB Spectral Energy Distribution (SED), shown in Fig.~1 (from
Hauser \& Dwek, 2001) shows the existence of a minimum between 3 and
10~$\mu m$ separating direct stellar radiation from the FIR part due to
radiation re-emitted by dust. The latter radiation contains at least a
comparable integrated power as the former, and perhaps as much as 2.5
times more. This ratio is much larger than what is measured locally
($\sim$30$\%$). The CIB is thus likely to be dominated by a population
of strongly evolving redshifted IR galaxies.  

\begin{figure}[t]
  \begin{center}
    \epsfig{file=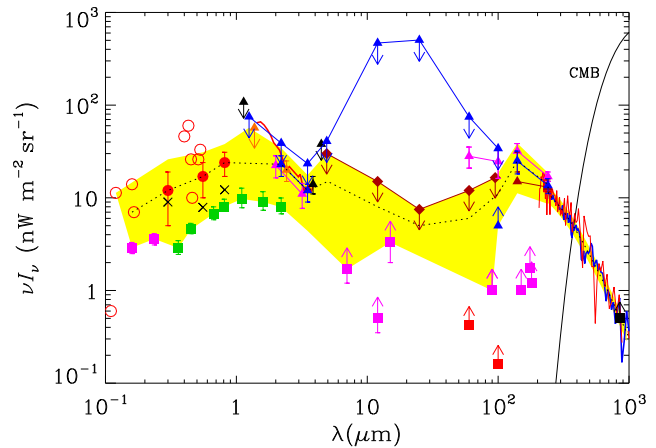, width=9cm}
  \end{center}
\caption{The Cosmic Infrared Background Spectral Energy Distribution
from Hauser \& Dwek (2001). This plot includes detection (solid line
in the submm part), upper limits (triangles and diamonds), and lower
limits from source counts (squares). The shaded region represents
current observational limits, and the solid rising curve in the mm
range represents the CMB..} 
\end{figure}

ISO revolutionized our view of the extragalactic universe with
sensitive cosmological surveys carried out in the mid- and far-
infrared. Of course, other cosmological surveys have been carried
out with SCUBA, MAMBO, Chandra, XMM, and in the optical range; all
give invaluable informations about galaxy evolution, but it is beyond
the scope of this short paper to review them.
Reviews of extragalactic results from ISO can be found in
Genzel \& Cesarsky (2000) and Franceschini et al. (2001), and a
comprehensive overview of ISO operations and science in Casoli et
al. (2000) (particularly in the Puget \& Guiderdoni paper). I will here
summarize the main results obtained in the MIR with ISOCAM (Sect.~2),
in the FIR with ISOPHOT (Sect.~3), and overview the models that ISO
data were able to constrain (Sect.~4). Sect.~5 emphasizes the
potential of the ISO archive, and in Sect.~6 I briefly explore the
future of IR (and mm) astronomy, in particular with SIRTF.

\section{MID INFRARED SURVEYS}
\subsection{Specificities}
The mid-infrared surveys have been conducted with ISO\-CAM (Cesarsky
et al., 1996), mainly at 7 and 15 $\mu m$; a few have been performed
at 4.5 $\mu m$ and 12 $\mu m$ (Clements et al., 1999).
At 7 $\mu m$ using the LW2 filter (e.g. Taniguchi et al., 1997; Oliver
et al., 2002), the contamination by stars is important and data at
other wavelengths (usually near infrared) are needed to identify
them. The decreasing k-correction is less favorable for detecting
higher redshift galaxies (dash line in Fig.~2, taken from Franceschini
et al., 2001). 

\begin{figure}[t]
  \begin{center}
    \epsfig{file=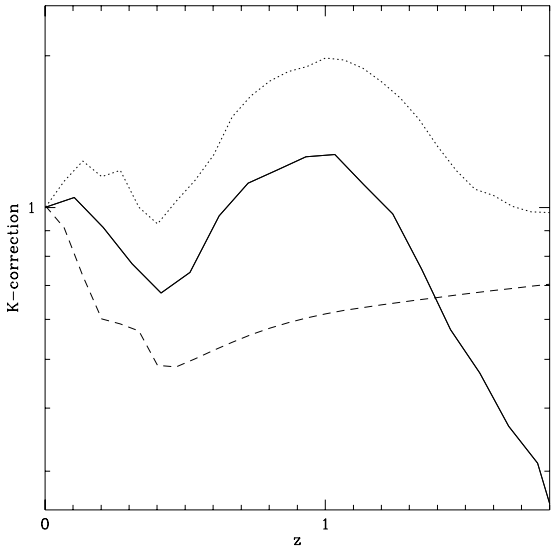, width=8cm}
  \end{center}
\caption{K-corrections from Francecshini et al (2001). Dash: M82
spectrum at 7 $\mu m$. Dot: inactive spiral at 15 $\mu m$. Line:
M82-like spectrum at 15 $\mu m$.}
\end{figure}

At 15 $\mu m$ using the LW3 filter (e.g Elbaz et al., 1999), the
stellar contamination is less of a problem and the k-correction of
galaxies is favorable for detecting sources up to redshifts around 1.4
(Fig.~2). Thus, a significant part of the
extragalactic results are based on 15 $\mu m$ observations.

\subsection{Source Counts}
Source counts at 15 $\mu m$ show an impressive consistency over
four decades of flux between surveys of different depths (e.g. Elbaz
et al., 1999, Gruppioni et al, 2002). Data come from large and shallow
surveys like \object{ELAIS} (e.g. Serjeant et al, 2000), to narrower and deeper
surveys (e.g. Serjeant et al, 1997, Aussel et al, 1999; Altieri et al,
1999). 
The number counts (summarized in Fig.~3 by Elbaz et al., 1999) are in
excess by a factor of 10, at the 0.5 mJy level, compared to non
evolution scenarios. The slope is steep ($\alpha = 3.0 \pm 0.1$ in the
differential counts, for $S_{\nu} > 0.4 $ mJy), and a turnover appears
at $S_{\nu} = 0.4 $ mJy. At fainter levels, the convergence (decrease)
is fast. 

\begin{figure}[t]
  \begin{center}
    \epsfig{file=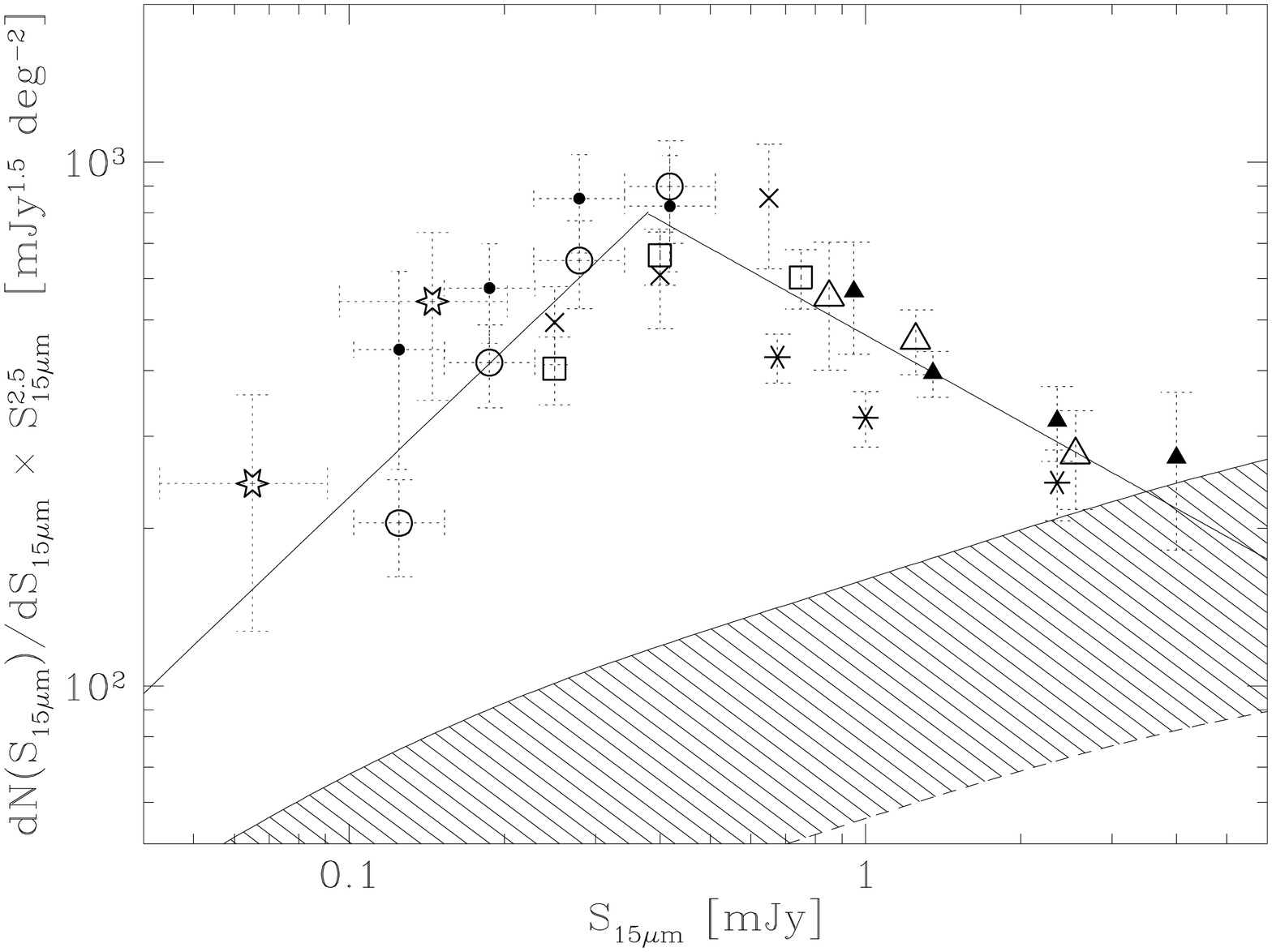, width=9cm}
  \end{center}
\caption{Differential source counts (normalized to Euclidean) at 15
$\mu m$ from Elbaz et al (1999). The shadowed area represents a
scenario with no evolution. Symbols represent different surveys.} 
\end{figure}

\subsection{Nature of the Galaxies}
With a median redshift of $z \sim 0.8$ and most of the sources with
$0.5 < z < 2$ (Aussel et al., 1999), the MIR sources have a mean
luminosity of $6 \times 10^{11} L_{\odot}$ (Elbaz et al., 2002). Most
of them are experiencing intense stellar formation of about 100
$M_{\odot}$yr$^{-1}$, which appears to be uncorrelated with the faint
blue galaxy population dominating the optical counts at z$\sim$0.7
(Ellis 1997; Elbaz et al. 1999). Fig.~4 (from Elbaz et al., 2002)
represents their HDF-N sample, and summarizes the properties of MIR
galaxies: redshift distribution, luminosity, and SFR (for 80\% of the
non-AGN galaxies). 

\begin{figure}[t]
  \begin{center}
    \epsfig{file=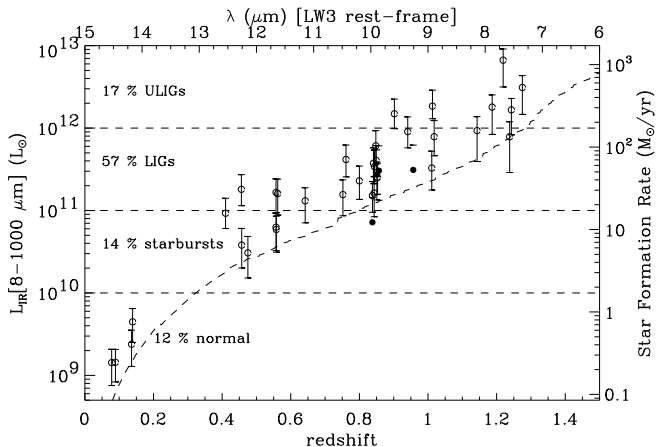, width=9cm}
  \end{center}
\caption{Infrared luminosity (left axis) and star formation rate (right
axis, for starbursts only) of 15 $\mu m$ $S>0.1$ mJy galaxies in the
HDF-N from Elbaz et al (2002). The dash curve represents the
sensitivity limit. This figure also illustrates the redshift
distribution of the sources, with a median redshift of $z=0.8$ (Aussel
et al., 1999).} 
\end{figure}

\subsection{Galaxy Evolution}
Elbaz et al (2002) present a summary of the galaxy evolution at 15
$\mu m$ and beyond. First, they show that 60\% of the CIB at 15 $\mu
m$ is created by LIRGs ($L > 10^{11} L_{\odot}$). Second, they show that
the comoving luminosity density at 15 $\mu m$ was about 55 times larger
at $z \sim 1$ than today. Third, still using some assumption about the
SED of the galaxies, they conclude that the comoving density of IR
luminosity radiated by dusty starbursts (which is directly related to
the star formation) was about $70 \pm 35$ times larger at $z \sim 1$
than today. At intermediate redshifts, Flores et al (1999) also
determined the star formation rate.
Finally, they show that about half of the CIB at 140 $\mu m$ is
produced by LIRGs and about one third by ULIRGs. Thus MIR galaxies
observed by ISOCAM have resolved about 75\% of the CIB at 140 $\mu
m$.

\section{FAR INFRARED SURVEYS}
\subsection{Specificities}
The far-infrared surveys have been conducted with ISO\-PHOT (Lemke et
al., 1996), mainly at 90 and 170 $\mu m$; a few have been performed at
60, 120, 150 and 180 $\mu m$ (Juvela et al., 2000; Linden-Voernle et al.,
2000). 
In the 50 to 100 $\mu m$ range, the C100 Ge:Ga camera is sensitive to the
peak of rest-frame emission from obscured star formation. 
The decreasing k-correction, in addition to a challenging data
processing, led to probe mainly the local universe. With 46 arcseconds
per pixel and a FWHM of about the same size, the angular resolution at
100 $\mu m$ is significant improved since IRAS.

In the 100 to 200 $\mu m$ range, the C200 stressed Ge:Ga camera is
sensitive to the very cool local galaxies as well as higher redshift
starburst galaxies, thanks to an advantageous k-correction
(e.g. Guiderdoni et al., 1997) and the behavior of the camera
(e.g. Lagache \& Dole, 2001). With 92 arcseconds per pixel and a FWHM
of about the same size, the angular resolution at 170 $\mu m$ is an
issue for source identifications.

\subsection{Source Counts}
Source counts at 170 $\mu m$ (e.g. Kawara et al., 1998; Puget et al.,
1999; Dole et al., 2001) exhibit a steep slope of $\alpha = 3.3 \pm 0.6$
between 180 and 500 mJy (lower left panel on Fig.~7) and, like in the
MIR range, show sources in 
excess by a factor of 10 compared with no evolution scenario. This
strong evolution was unexpected and quite difficult to reproduce with
models. Matsuhara et al (2002) extended the source count analysis to
fainter fluxes and still detected a strong evolution.

At 60 and 90 $\mu m$, the situation is less clear, but preliminary source
counts (Kawara et al., 1998; Efstathiou et al., 2000; Linden-Voernle et
al., 2000) are compatible with no evolution scenarios on their bright
end, and begin to show evolutionary effects on their bright end. New
processing techniques (see Sect.~5) will certainly allow to go deeper.

\begin{figure}[t]
  \begin{center}
    \epsfig{file=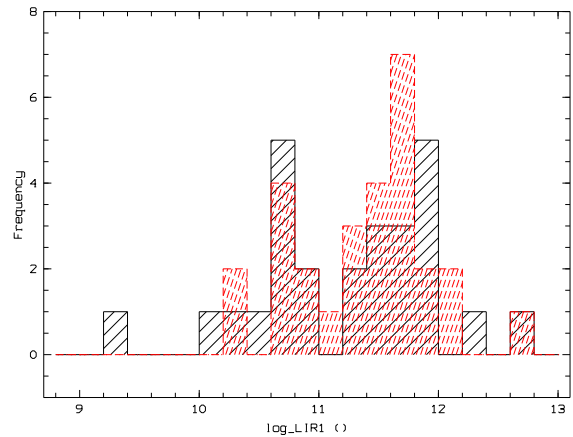, width=9cm}
  \end{center}
\caption{Distribution of IR luminosities from Patris et al
(2002). Solid line: FIRBACK FSM 170 $\mu m$ sample ($S>200$ mJy); Dash:
IRAS 100 $\mu m$ sample from Bertin et al (1997) with $110 < f_{60} <
200$ mJy.} 
\end{figure}

\subsection{Nature of the Galaxies}
Determining the nature of the FIR galaxies has been a longer process
than in the MIR, mainly because of the difficulty to find the shorter
wavelength counterparts in a large beam. Various techniques
have been used to overcome this problem, one of the most successful
being the identification using 20cm radio data (e.g. Ciliegi et al.,
1999). Another technique is the FIR multiwavelength approach (Juvela et
al., 2000) that helps constraining the position and the SED; it
also helps to separate the cirrus structures from the extragalactic
sources. A variation is to use ISOCAM and ISOPHOT data, like the ELAIS
Survey (Oliver et al., 2000; Serjeant et al., 2000,2001). 
Finally, the Serendipity Survey (Stickel et al., 1998, 2000), by covering
large and shallow areas, allows to detect many bright objects easier
to follow-up or already known.

FIR ISO galaxies can be sorted schematically into two populations.
First, the low redshift sources, typically $z<0.3$ (e.g Serjeant et
al., 2001; Patris et al., 2002, Kakazu et al., 2002), have moderate IR
luminosities, below $10^{11} L_{\odot}$, and are cold (Stickel et al.,
2000). Second, sources at higher redshift, $z \sim 0.3$ (Patris et
al., 2002, see Fig.~5) and beyond, $z \sim 0.9$ (Chapman et al., 2002)
are more luminous, typically $L > 10^{11} L_{\odot}$, and appear to be
cold. Serjeant et al. (2001) derived the Luminosity Function at 90 $\mu
m$, and started to detect an evolution compared to the local IRAS 100 $\mu
m$ sample.

\subsection{Fluctuations of the CIB}
Sources below the detection limit of a survey create fluctuations. If
the detection limit does not allow to resolve the sources dominating
the CIB intensity, which is the case in the FIR with ISO,
characterizing these fluctuations gives very interesting information
on the spatial correlations of these unresolved sources of
cosmological significance. An example of the modeled redshift distribution of
the unresolved sources at 170 $\mu m$ can be found in Fig.~12 of
Lagache et al (2002); the sources dominating the CIB fluctuations have
a redshift distribution peaking at $z \sim 0.9$.
After the pioneering work of Herbstmeierer al (1998) with ISOPHOT,
Lagache \& Puget (2002) discovered them at 170 $\mu m$ in the \object{FIRBACK}
data, followed by other works at 170 and 90 $\mu m$ (Matsuhara et al.,
2000; Puget \& Lagache, 2000; Kiss et al, 2001). Fig.~6 shows the CIB
fluctuations in the FN2 field by Puget \& Lagache, (2000), at
wavenumbers $0.07 < k < 0.4$ arcmin$^{-1}$.

\begin{figure}[t]
  \begin{center}
    \epsfig{file=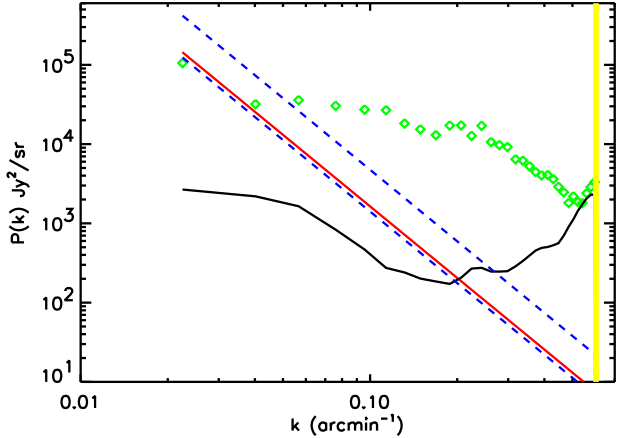, width=9cm}
  \end{center}
\caption{Fluctuations of the CIB in a power spectrum analysis of the
FIRBACK/ELAIS N2 field at 170 microns by Puget \& Lagache
(2000). Observed power spectrum: diamond; straight continuous line:
the best fit cirrus power spectrum; dash line: cirrus power spectrum
deduced from Miville-Desch\^enes et al (2002); continuous curve:
detector noise.} 
\end{figure}

\section{MODELS}
ISO provided high quality data, taken in the frame of the cosmological
survey programs, as well as nearby galaxies or serendipitous
programs. 
The first observable to be widely used as a constraint on the models
is the source counts at 15 and 170 $\mu m$, but also at 7 and 90 $\mu
m$. The redshift distributions, mainly at 15 $\mu m$ (and soon at 170
$\mu m$) put additional constraints, as well as a better knowledge of
the nature of the galaxies: AGN vs Starburst, luminosities, star
formation rate. The global star formation rate, as derived in
part from ISOCAM data, adds other constraints to the models. 
Finally, a few models use explicitly the constraint of the level of
the CIB fluctuations in the far infrared.

These observables and constraints induced a burst in the development
of new models in the late nineties, among which: 
Roche \& Eales (1999),
Tan et al. (1999),
Devriendt \& Guiderdoni (2000),
Dole et al. (2000),
Wang (2000),
Chary \& Elbaz (2001),
Franceschini et al. (2001),
Malkan \& Stecker (2001),
Pearson (2001),
Rowan-Robinson (2001),
Takeuchi (2001),
Xu et al. (2001),
Balland et al. (2002),
Lagache et al. (2002),
Totani \& Takeuchi (2002),
Wang (2002).
This development was also made possible because of the better
knowledge of the galaxies SED in the infrared (e.g. Dale et al., 2000;
Helou et al., 2000; Lutz et al., 2000; Stickel et al., 2000, Tuffs \&
Popescu, 2002), with the availability of the SED of the CIB (Gispert
et al., 2000; Hauser \& Dwek, 2001), and of complementary data sets
(X-rays, optical, NIR, submm, radio).

Even if there is no unique scenario reproducing all the observables,
the models help now to have a more coherent picture of galaxy evolution. 
It is beyond the scope of this paper to review the models and their
predictions, we thus give here only two examples.
Fig.~7 shows the observed source counts at 15, 60, 170 and 850 $\mu m$
and the fits by Lagache et al (2002). Using other kind but similar
evolutions (pure density and pure luminosity evolutions in addition to
the density+luminosity evolution), Chary \& Elbaz (2001)
predict the star formation rate and compare it with data and Xu et al
(2001) (Fig.~8). Combined with the different approach of Gispert et al
(2000), the SFR is well constrained in the redshift range $1 < z < 3$.

\begin{figure}[ht]
  \begin{center}
    \epsfig{file=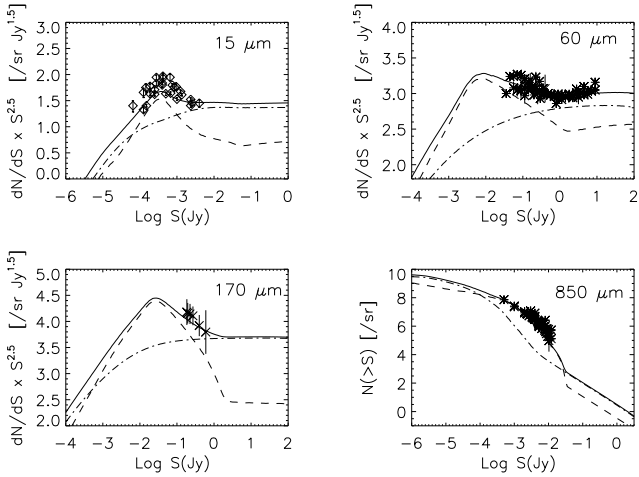, width=9.2cm}
  \end{center}
\caption{Observed source counts at 15, 60, 170 and 850 $\mu$m with the
model of Lagache et al (2002). Symbols represent data (CAM at 15 $\mu
m$, IRAS at 60 $\mu m$, PHOT at 170 $\mu m$ and SCUBA at 850 $\mu
m$. Solid line: model prediction. Dash: LIRG/starburst
component. Dot-dash: normal/cold galaxy component.} 
\end{figure}

\begin{figure}[ht]
  \begin{center}
    \epsfig{file=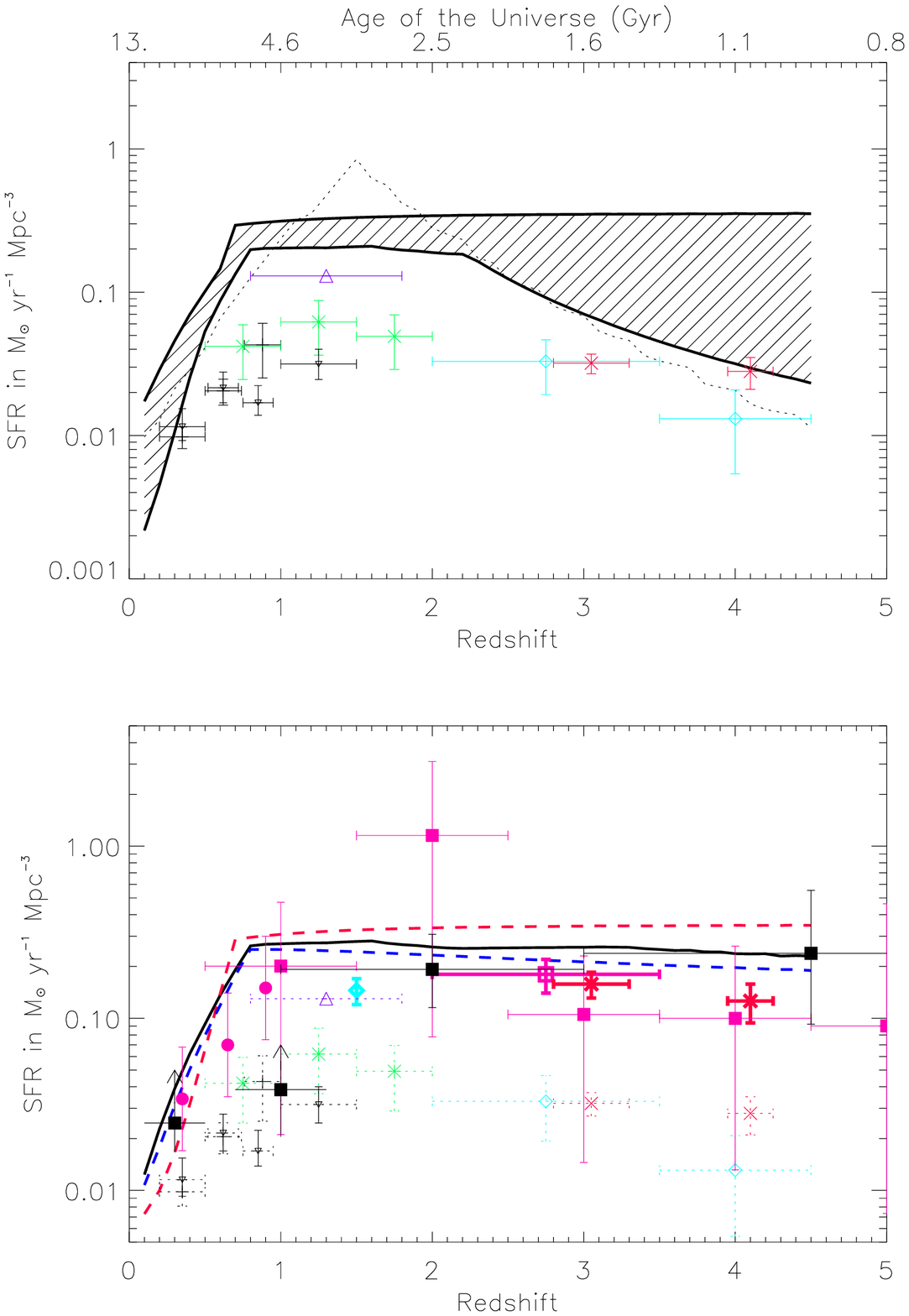, width=8.7cm}
  \end{center}
\caption{Star Formation Rate from Chary \& Elbaz (2001). Upper panel:
min and max range from their model, and observed UV/opt data; dots
represent Xu et al (2001) model.
Lower panel: 3 different evolution scenarios from their model and data
corrected for extinction. Line: pure luminosity; upper dash: pure
density; lower dash: luminosity+density.}
\end{figure}

\section{POTENTIAL OF ISO DATA}
Even if the ISO data led to an impressive scientific return with a
high efficiency (about 62\% of the data are published) in the field of
the cosmological evolution of the galaxies, many other results are
still to come, based on the amount of unpublished data available in
the archive.

In addition, new techniques of data reduction are now available,
thanks to a better understanding of detector behaviors, both
for the Silicon Array (CAM)(Coulais \& Abergel, 2000;
Miville-Desch\^enes et al., 2000; Lari et al., 2001, 2002; Vaccari, 2002)
and the Germanium detectors (PHOT) (Coulais et al., 2000).
Processing unpublished data and/or reprocessing published data
allows us to reach better sensitivity levels with a higher confidence
(e.g. for ISOCAM HDF-N: Serjeant et al., 1997; Aussel et al., 1999;
D\'esert et al., 1999. E.g. for ISOPHOT Lockman Hole: Kawara at al.,
1998; Rodighiero et al., 2002). 

Concerning the CIB and the cirrus foreground, a brilliant example of
what can be done from the archive is the work of Kiss et al
(2001,2002), detailed in this conference.
Furthermore, the ISO data may induce discoveries in other data sets. A
recent example concerns the fluctuations of the CIB in the
far infrared. Lagache \& Puget (2000) and Puget \& Lagache (2000)
measured the level of the fluctuations of the CIB at $170 \mu m$ in
the FIRBACK fields. Given the SED of the CIB, it 
was possible to predict the fluctuation level at 60 and 100 $\mu m$
(e.g. Lagache, Dole, Puget, 2002), and show that this might be
detectable in the IRAS data. 
As a consequence, Miville-Desch\^enes et al. (2002) recently reported
the first detection of the CIB fluctuations at 60 and 100 $\mu m$ in
the IRAS data. 

\section{THE NEXT STEPS: SIRTF AND BEYOND}
SIRTF, the Space Infrared Telescope Facility, to be laun\-ched in early
2003, is expected to improve our view of the IR universe over the next five
years, thanks to the recent technological developments in IR detector
arrays. A number of multiwavelength cosmological surveys are scheduled with MIPS
(Multiband Imaging Photometer for SIRTF) at 24, 70 and 160 $\mu m$
(Dole et al., 2003\footnote{see also http://lully.as.arizona.edu}) and
IRAC (Infrared Array Camera) at 3.6, 4.5, 5.8 and 8.0 $\mu m$.
In the frame of the Guaranteed Time (GTO) or Legacy (GOODS and SWIRE)
Surveys, about 80 Sq. Deg. will be observed at different depths.
Probing the properties of the galaxies with samples of statistical
significance up to redshifts of 2.5 or more will be possible. Fig.~9
(from Dole et al, 2003) shows predictions of the redshift distribution
at 24 $\mu m$ for MIPS surveys. 

The next steps after ISO and SIRTF for understanding the
infrared and submillimeter Universe from space will be ISAS's ASTRO-F
in 2004, ESA's Herschel \& Planck (around 2007), and NASA's NGST
(around 2010+), in addition to the ground-based ALMA (from 2005). These
telescopes are the promise for a lot of exciting science to be done (at
least) in the next 15 years ! 

\begin{figure}[t]
  \begin{center}
    \epsfig{file=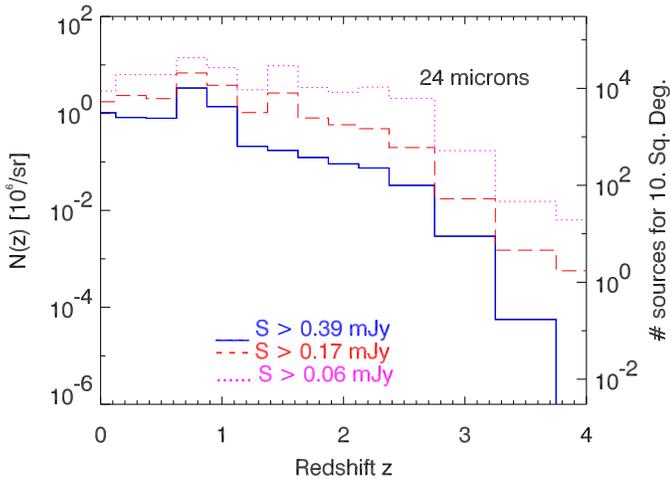, width=8.5cm}
  \end{center}
\caption{Predicted redshift distribution of sources observed at 24
$\mu m$ with MIPS on SIRTF, from Dole et al (2003), for various
$5 \sigma$ surveys depths.
Line: GTO shallow survey or SWIRE. Dash: GTO deep. Dot:
GTO ultradeep or GOODS.}
\end{figure}

\begin{acknowledgements}
I warmly acknowledge the organizers for having invited me to give this
review at this very interesting and enjoyable conference (especially
for the tapas, the hot chocolate around the San Juan bonfire, and the
cultural wonders).
I also acknowledge the Programme National de Cosmologie and the Centre
National d'Etudes Spatiales (CNES) for travel support for the work
with my collaborators in France, and the funding from the MIPS
project, which is supported by NASA through the Jet Propulsion
Laboratory, subcontract \# P435236. 
\end{acknowledgements}


\end{document}